\documentstyle[11pt,appb,epsfig]{article}
\newcommand{\be}{\begin{equation}}
\newcommand{\ee}{\end{equation}}
\newcommand{\bea}{\begin{eqnarray}}
\newcommand{\eea}{\end{eqnarray}}

\begin{document}
\title{PERSPECTIVES OF SCALAR- AND VECTOR-MESON PRODUCTION IN
HADRON-NUCLEUS REACTIONS\footnote{Supported by BMBF, FZ J\"ulich
and the Heisenberg-Landau program }}
\author{W. CASSING\thanks{In collaboration with E. L. Bratkovskaya, M. B\"uscher,
Ye. S. Golubeva, V. Grishina, V. Hejny, L. A. Kondratyuk, G. I.
Lykasov, M. V. Rzjanin, A. Sibirtsev, H. Str\"oher}
\\ Institut f\"ur Theoretische Physik, Universit\"at Giessen \\
D-35392 Giessen, Germany}
\date{ }
\maketitle

\begin{abstract}
The production and decay of vector mesons ($\rho, \omega$) in $pA$
reactions at COSY energies is studied with particular emphasis on
their in-medium spectral functions. It is explored within
transport calculations, if hadronic in-medium decays like
$\pi^+\pi^-$ or $\pi^0 \gamma$ might provide complementary
information to their dilepton ($e^+e^-$) decays. Whereas the
$\pi^+ \pi^-$ signal from the $\rho$-meson  is found to be
strongly distorted by pion rescattering, the $\omega$-meson Dalitz
decay to $\pi^0 \gamma$ appears promising even for more heavy
nuclei. The perspectives of scalar meson ($f_0, a_0$) production
in $pp$ reactions are investigated within a boson-exchange model
indicating that the $f_0$-meson might hardly be detected in these
collisions in the $K \bar{K}$ or $\pi \pi$ decay channels whereas
the exclusive channel $pp \rightarrow d a_0^+$ looks very
promising.
\end{abstract}
\maketitle

 PACS numbers: 13.25.-k; 13.60.Le; 13.75.-n

\section{Introduction}
The modification of the vector meson properties~\cite{Medium} --
\cite{Medium6} in nuclear matter has become a challenging subject
in dilepton physics from $\pi^- A$, $p A$ and $A A$ collisions.
Here the dilepton ($e^+e^-$) radiation from $\rho$'s and
$\omega$'s propagating in finite density nuclear matter is
directly proportional to their spectral function which becomes
distorted in the medium due to the interactions with
nucleons~\cite{Propagation} -- \cite{m6}. Apart from the vacuum
width $\Gamma^0_V \ (V= \rho, \omega$) these modifications are
described by the real and imaginary part of the retarded  self
energies $\Sigma_V$, where the real part $\Re \Sigma_V$ yields a
shift of the meson mass pole and the additional imaginary part
$\Im \Sigma_V$ (half) the collisional broadening of the vector
meson in the medium. We recall that the meson self energy in the
$t-\rho$ approximation is proportional to the complex forward $V
N$ scattering amplitude $f_{VN}(P,0)$ and the nuclear density
$\rho(X)$, i.e. $ \Sigma_V(P,X) = - 4 \pi \rho(X) f_{V N}(P,0)$.
The scattering amplitude itself, furthermore, obeys dispersion
relations between the real and imaginary parts
\cite{Dispersion,d2,d3} while the imaginary part can be determined
from the total $V N$ cross section according to the optical
theorem. Thus the vector meson spectral function
\begin{eqnarray}
A_V(X,\!P){=}\frac{\Gamma_V (X,\!P)}{(P^2{-}
M_{V}^{2}{-}\Re\Sigma_V(X,\!P))^{2}{+} \Gamma_V (X,\!P)^{2}/4} ,
 \label{spectral}
\end{eqnarray}
where $\Gamma_V(X,P){=}{-}2\Im\Sigma_V(X,P)$, can be constructed
once the $V N$ elastic and inelastic cross sections are known.
Note that in (\ref{spectral}) all quantities depend on space-time
$X$ and four-momentum $P$.

\section{Production and decay of vector mesons at finite density}

As mentioned before, the in-medium vector meson spectral functions
can be measured directly by the leptonic decay $V {\to}e^+e^-$ or
the Dalitz decay $\omega{\to}\pi^0\gamma$, respectively. The
vector meson production in $pA$ collisions can be considered as a
natural way~\cite{Recent} to study the $\rho$- and
$\omega$-properties at normal nuclear density under rather well
controlled conditions. This also holds for the photo-production of
vector mesons on nuclei \cite{Mosel}.

The following calculations have been performed for $p{+}A$
collisions at 2.4--2.5~GeV by introducing a real and imaginary
part of the vector meson self energy as
\begin{equation}
\label{poten}
 U_V = \frac{\Re \Sigma_V}{2 M_0} \simeq M_0 \
\beta \frac{ \rho(X)}{\rho_0} \end{equation} and width
\begin{equation}
\label{width} \Gamma^\ast = \frac{\Im \Sigma_V}{2 M_0} \simeq
\Gamma_V^0{+}\Gamma_{coll} \frac{\rho(X)}{\rho_0}
\end{equation}
in the $t{-}\rho$-approximation. Here $M_0$ and $\Gamma_V^0$
denote the bare mass and width of the vector meson in vacuum while
$\rho(X)$ is the local baryon density and $\rho_0$=0.16~fm$^{-3}$.
The parameter $\beta{\simeq}$--0.16 was adopted from the models in
Refs.~\cite{Medium} -- \cite{Medium6}. The predictions for the
$\omega$-meson collisional width $\Gamma_{coll}$ at density
$\rho_0$ range from 20 to 50~MeV~\cite{Medium,Medium1,OmegaN} --
depending on the number of $\omega{N}$ final channels taken into
account --  while the collisional width of the $\rho$-meson should
be about 100 -- 120 MeV at $\rho_0$ due to the strong coupling to
baryon resonances (cf. Fig. 6 of Ref. \cite{d3}).

\subsection{$\rho^0$-mesons}
 In Ref. \cite{Model} the $\pi^+ \pi^-$ decays of
$\rho^0$-mesons produced in $p A$ reactions has been investigated
with the aim of testing the in-medium $\rho^0$-spectral function
at normal nuclear matter density.

\begin{figure}
\psfig{file=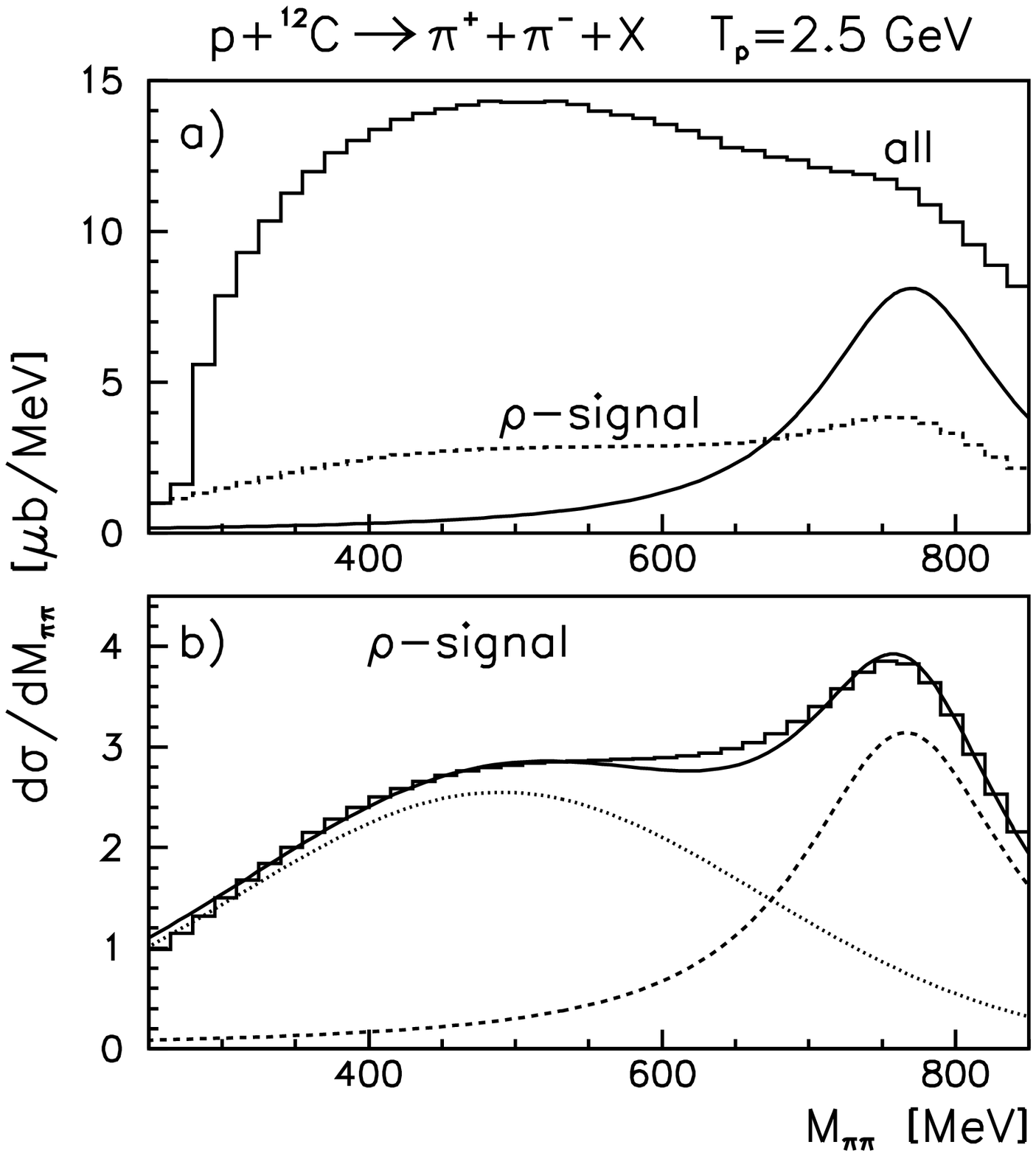,width=6cm} \hskip 0.3cm
\psfig{file=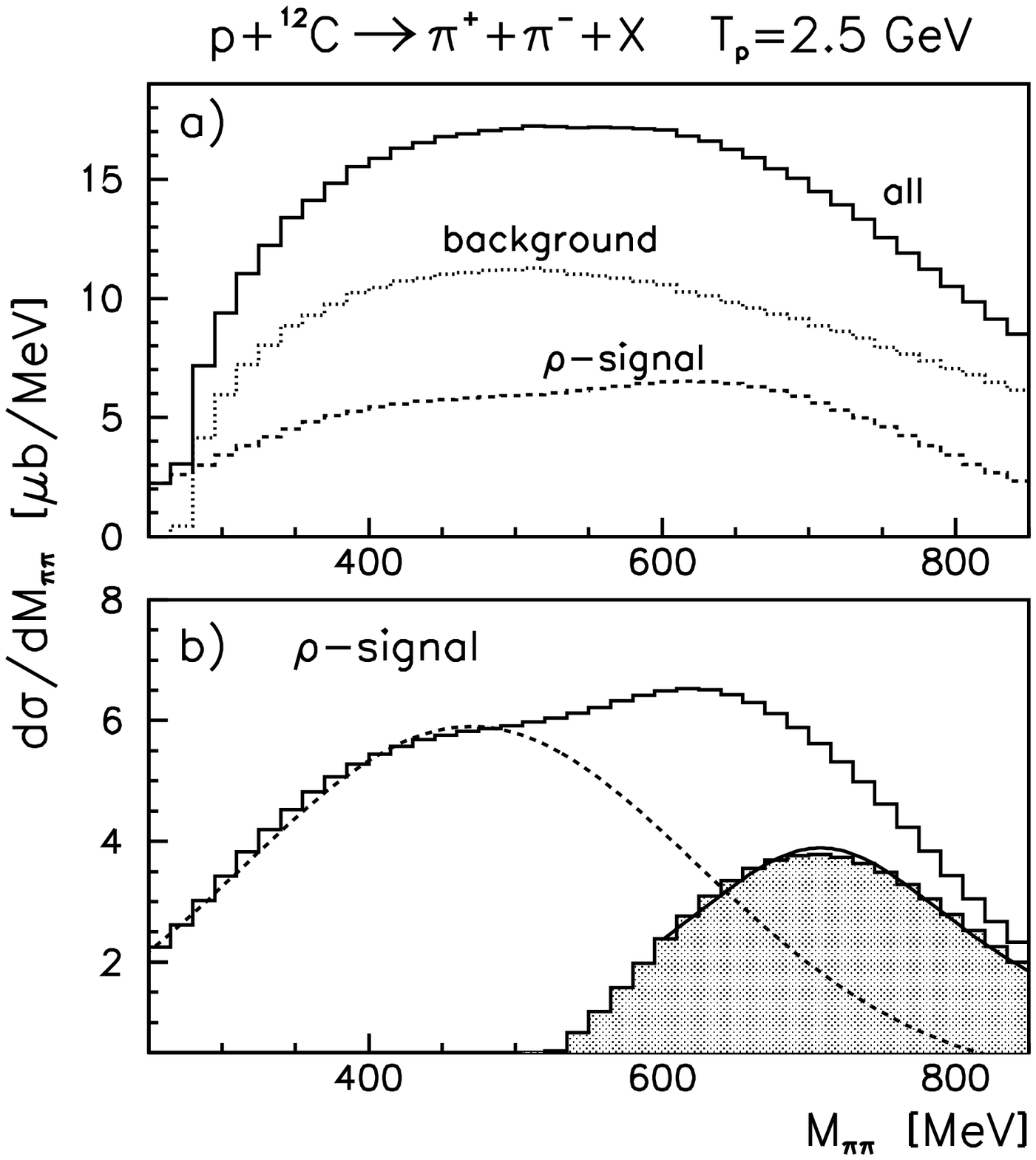,width=6cm} {\small \caption[]{\label{cbuu3}
(l.h.s.) The invariant mass distribution of two pions from
$p+{^{12}C}$ collisions at 2.5 GeV using the bare spectral
function of the $\rho$-meson for $p_z \leq$ 2 GeV/c and $p_T \leq$
0.8 GeV/c. Displayed are the total $\pi^+\pi^-$ invariant mass
spectrum (all), the $\rho$ signal after subtraction of the
combinatorial background (dashed line) as well as a Breit-Wigner
distribution with the bare $\rho$-meson properties normalized to
the total $\rho$-production cross section from the transport
approach (see text).
 \\ (r.h.s.) The invariant mass distribution of
two pions from $p+{^{12}C}$  at 2.5 GeV ($p_z \leq$ 2 GeV/c and
$p_T \leq$ 0.8 GeV/c) employing the in-medium modification of the
$\rho$-meson according to Eq. (2) for $\beta$ = - 0.16.  In b) the
solid histogram shows the $\rho$-meson signal from a); the dashed
line is a statistical fit while the hatched histogram is the
difference between the $\rho$ signal and the statistical fit,
 which can be described again by a
Breit-Wigner function (solid line) with $M_0^*$ = 708 MeV and
$\Gamma$ = 270 MeV (see text).}}
\end{figure}

Some results of the transport model studies from Ref. \cite{Model}
are displayed in Fig. \ref{cbuu3} for the reaction $p + ^{12}C$ at
2.5 GeV. The l.h.s. displays the invariant mass distribution of
two pions  using the bare spectral function of the $\rho$-meson
for $p_z \leq$ 2 GeV/c and $p_T \leq$ 0.8 GeV/c. In a) the solid
histogram shows the total $\pi^+\pi^-$ invariant mass spectrum
while the dashed histogram indicates the $\rho$ signal after
subtraction of the combinatorial background. The solid line is a
Breit-Wigner distribution with the bare $\rho$-meson properties
normalized to the total $\rho$-production cross section from the
transport approach as expected in case of no final state
interactions.  In b) the histogram shows the $\rho$-meson signal
from a) while the dotted line is the "statistical spectrum"; the
dashed line is a Breit-Wigner function with the $\rho$-meson
properties $M_0=766.9$~MeV and ${\Gamma}_0=173$~MeV while the
solid line shows the sum. The dashed histogram in the r.h.s. of
Fig. \ref{cbuu3}a) represents the $\rho$-signal as obtained by
background subtraction from uncorrelated $\pi^-\pi^-$ pairs in
case of the in-medium modification (2) with $\beta$ = -0.16. It is
seen that not only the shape of the mass spectra, but also the
absolute normalization are different from the calculations with
the bare $\rho$-properties (l.h.s.). The $\rho$-meson signal is
shown again in Fig.\ref{cbuu3}b) (r.h.s.) in terms of the solid
histogram, which at first glance is quite complicated to analyze
in order to extract in-medium properties of the $\rho$-meson. To
this aim we have fitted the spectrum with a statistical
distribution \cite{Model} as shown in Fig.\ref{cbuu3}b) by the
dashed line. The hatched histogram in Fig.\ref{cbuu3}b) then shows
the difference between the $\rho$-mass spectrum and the
statistical distribution; it can  be fitted by a Breit-Wigner
distribution with $M_0^*=$708~MeV and ${\Gamma}$=270~MeV.

In comparison with the l.h.s. of Fig.\ref{cbuu3}b) a dropping of
the $\rho$-meson mass thus might be extracted from the invariant
mass distribution of two pions for light nuclei such as $^{12}C$.
However, the analysis is partly model dependent and the experiment
requires large area detectors.

\subsection{$\omega$-mesons}
The situation changes for the in-medium $\omega$-meson Dalitz
decay since here only a single pion might rescatter whereas the
photon escapes practically without reinteraction. Again the
calculations are performed within the transport model used before
for $\rho^0$ and $K^\pm$ studies~\cite{d2,Model} by taking into
account both the direct $pN{\to}{\omega}pN$,
$pN{\to}{\omega}pN\pi$ and secondary $pN{\to}{\pi}NN$,
${\pi}N{\to}{\omega}N$ production mechanisms employing the cross
sections from Ref.~\cite{Elementary} as well as ${\omega}N$
elastic and inelastic interactions in the target nucleus (with
cross sections from Ref.~\cite{OmegaN}) and accounting for $\pi^0
N$ rescattering. Here the $\omega$ propagation is described by the
Hamilton's equation of motion and its Dalitz decay to $\pi^0
\gamma$ by Monte Carlo according to the survival probability
$P_\omega(t{-}t_0){=} \exp({-}\Gamma_V^0(t{-}t_0))$ in the rest
frame of the meson that was created at time $t_0$.

\begin{figure}[h]
\vspace{-15mm} \psfig{file=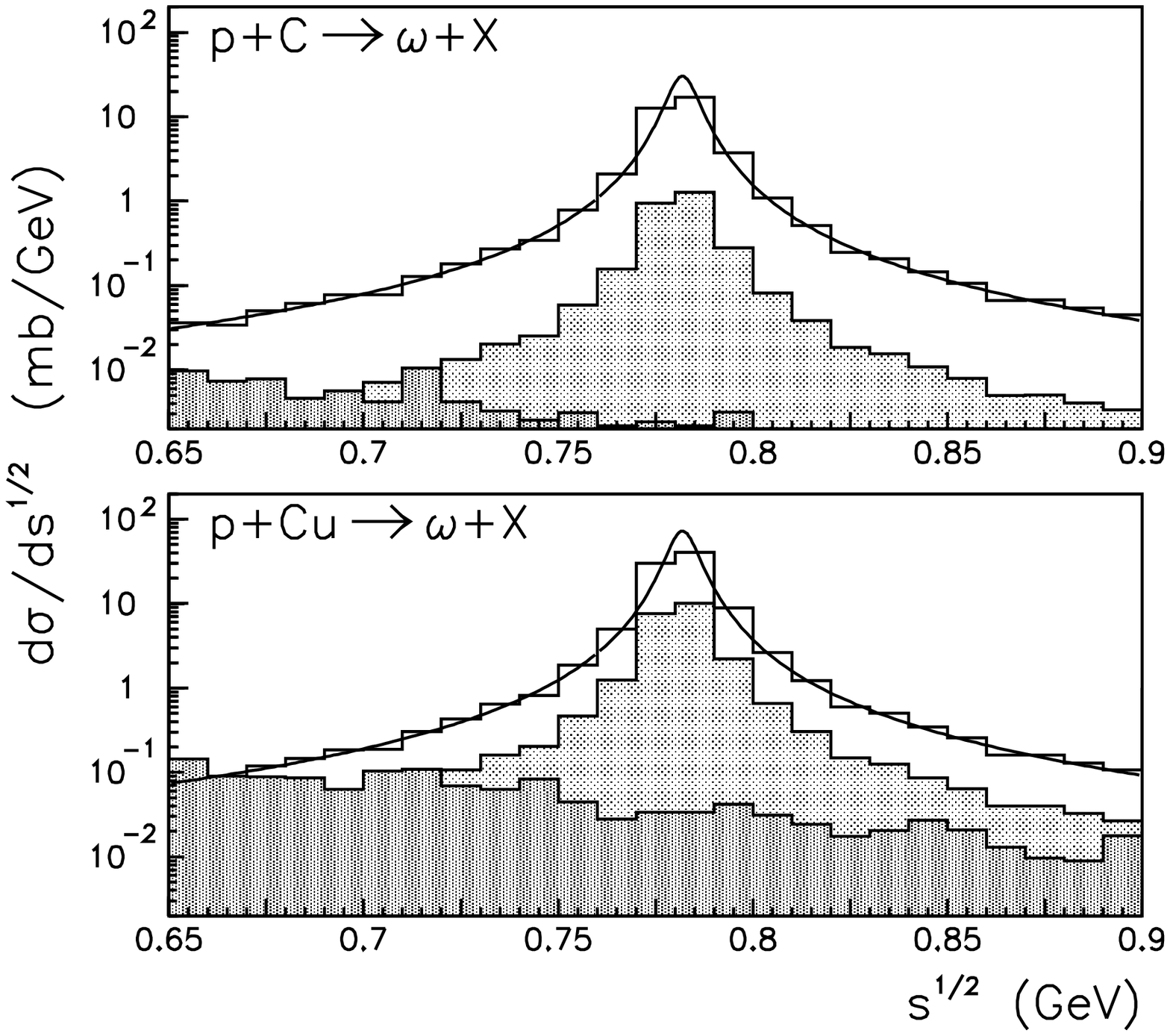,width=6.2cm,height=8.cm}
\hskip 0.2cm \psfig{file=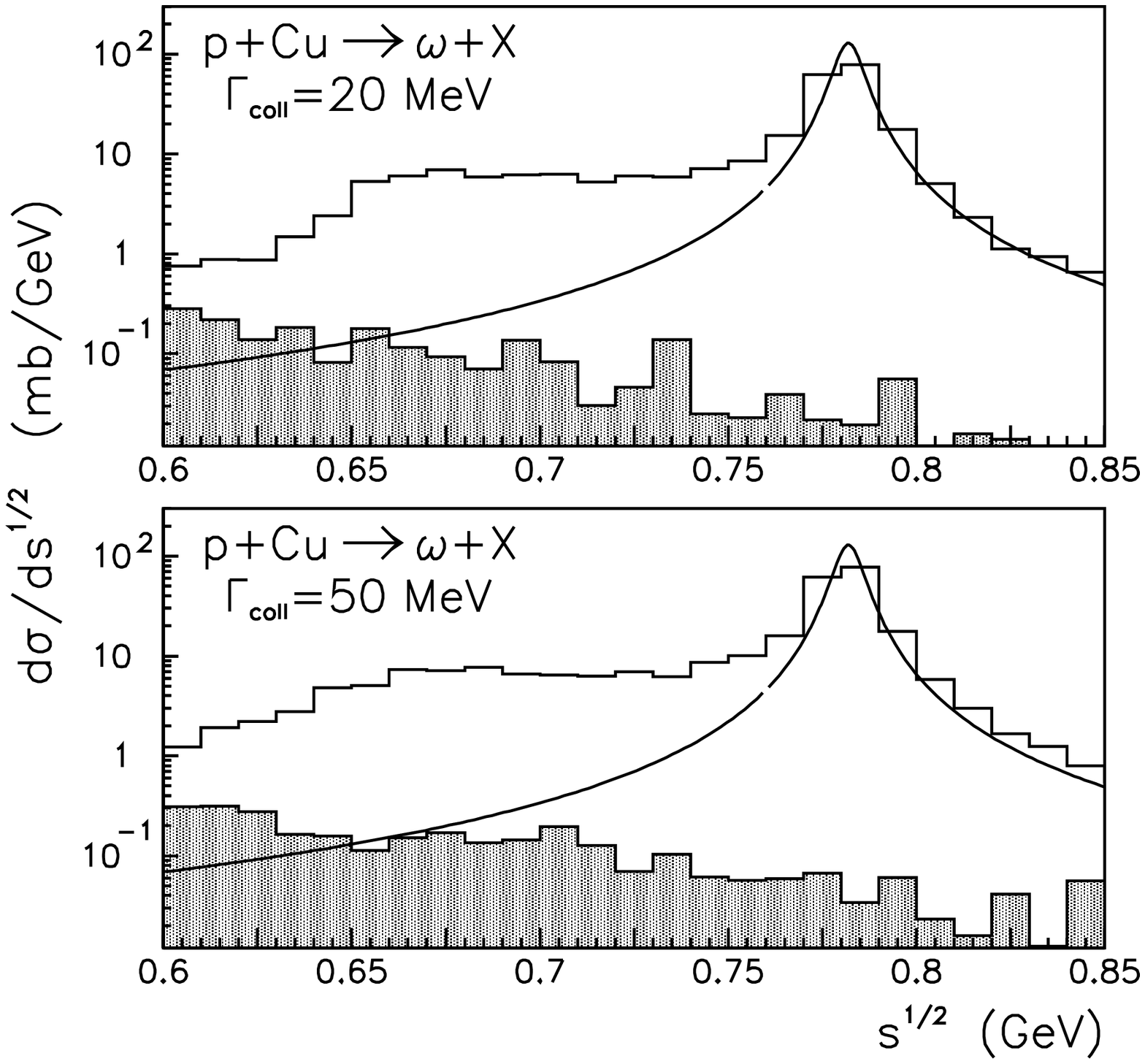,width=6.cm,height=7.6cm}
 \vspace{1mm} {\small
\caption{(l.h.s.) The $\pi^0\gamma$ invariant
mass spectra from $p{+}C$ and $p{+}Cu$ collisions at a beam energy
of 2.4~GeV calculated without in-medium modifications of the
$\omega$-meson. The different contributions are explained in the
text.  (r.h.s.) The $\pi^0\gamma$ invariant mass spectra from
$p{+}Cu$ collisions at 2.4~GeV calculated for $\Gamma_{coll}$=20
and 50~MeV (open histograms) when employing the potential
(\ref{poten}) for $\beta$ = -0.16. The hatched histograms indicate
the contributions from $\omega$-mesons, that decay at finite
density and include $\pi^0$ rescattering, while the solid lines
show the $\omega$-spectral function in vacuum for comparison.}}
\label{ankeom6}
\end{figure}

We  first present the results without employing any medium
modifications for the $\omega$-meson. The l.h.s. of
Fig.~\ref{ankeom6} shows the resulting $\pi^0\gamma$ invariant
mass spectrum for $p{+}C$ and $p{+}Cu$ collisions at a beam energy
of 2.4~GeV.  The solid histogram  displays the spectral function
of $\omega$'s, which decay outside the nucleus at densities
$\rho{\le}$0.05~fm$^{-3}$; the distribution from the
$\omega{\to}\pi^0\gamma$ decay for $\rho{>}$0.05~fm$^{-3}$ for
events  without $\pi^0$-rescattering is shown by the light hatched
areas while events that involve ${\pi^0}N$ elastic or inelastic
scattering are displayed in terms of the dark areas. The solid
lines show the Breit-Wigner distribution determined by the pole
mass and width for the free $\omega$-meson. As can be seen from
Fig.~\ref{ankeom6} (l.h.s.) most of the $\omega$-mesons from the
$^{12}C$ target decay in the vacuum (92\%) and consequently
$\omega$ decays in the medium as well as $\pi^0$ rescattering are
rather scarce. The situation changes for a $Cu$ target where
$\pi^0 \gamma$ coincidences from finite density are more frequent
(19\% $\omega$'s decay inside the target), however, also $\pi^0$
rescattering gives a substantial background which in the invariant
mass range of interest can approximately be described by an
exponential tail in $\sqrt{s} = M_\omega$. Note, that
experimentally the in-medium $\omega$ spectral function can only
be observed by the $\pi^0\gamma$ invariant mass distribution from
$\omega$-mesons decaying inside the nucleus without
$\pi^0$-rescattering.

Now we examine the feasibility of a direct detection of an
in-medium modification of the $\omega$-spectral function via the
$\pi^0\gamma$ invariant mass spectrum. The solid histograms in
Fig.~\ref{ankeom6} (r.h.s.) show the calculated $\pi^0\gamma$
invariant mass spectra from $p{+}Cu$ collisions at 2.4~GeV
calculated with $\Gamma_{coll}$=20~MeV (upper part) and
$\Gamma_{coll}$=50~MeV (lower part) while employing the
potential~(\ref{poten}) with $\beta$ = -0.16. The results are
shown for an 'experimental' mass resolution of 10~MeV. The solid
lines indicate the Breit-Wigner distribution given by the mass and
width for the vacuum $\omega$-meson for comparison. The results
indicate a substantial enhancement of the low mass $\pi^0\gamma$
spectra due to the contribution from $\omega$-mesons decaying
inside the nucleus while feeling the attractive
potential~(\ref{poten}). Apparently, there is no substantial
difference between the $\pi^0\gamma$ spectra calculated with
$\Gamma_{coll}$=20 and 50~MeV since the shape of the low invariant
mass spectrum is dominated by the (density dependent) in-medium
shift of the $\omega$-pole. Only above the vacuum $\omega$-pole
mass one can see a slightly enhanced yield in case of the higher
collisional broadening. Moreover, the contribution from
'distorted' $\omega$-mesons (due to $\pi^0$ rescattering), which
is shown in Fig.~\ref{ankeom6} by the dark hatched histograms, is
small and represents an approximately exponential background in
the available $\pi^0\gamma$ energy.

As demonstrated in Ref. \cite{Hejny} $\omega$-meson mass shifts
comparable to Fig. 2 (r.h.s.) may also be observed for targets as
light as $^{12}C$ or heavy as $^{208}Pb$. Such experiments might
be carried out at COSY with neutral particle detectors looking for
the $3\gamma$ invariant mass distribution and gating on events
where 2 $\gamma$'s yield the invariant mass of a $\pi^0$. This
program is complementary to dilepton studies in $pA$ reactions
with the HADES detector at GSI Darmstadt~\cite{Hades}.

\section{Scalar meson production}

The scalar meson sector plays a very important role in the physics
of hadrons. Nevertheless, the structure of the lightest scalar
mesons  a$_0(980)$ and f$_0(980)$ is not yet understood and is one
of the most important  topics of hadronic physics
(\cite{Hadron99a,Hadron99b} and references therein). It has been
discussed that they could be either "
$\mathrm{q}\overline{\mathrm{q}}$ states", "Four-quark
cryptoexotic states", $\mathrm{K}\overline{\mathrm{K}}$ molecules
or  vacuum scalars (\cite{Hadron99a}). Nowadays, theory gives some
preference to the Unitarized Quark Model (UQM) proposed by
Tornqvist \cite{Tornqvist}, however, other options cannot be ruled
out experimentally. Moreover, there is a strong mixing between the
uncharged $a_0(980)$ and the $f_0(980)$ due to a coupling to
$\mathrm{K}\overline{\mathrm{K}}$ intermediate states. It is,
therefore, important to study independently the uncharged and
charged components of the a$_0(980)$ because the latter are not
mixed with the f$_0(980)$ and preserve their original quark
content.

\subsection{The reaction $NN{\to}f_0NN$}
The production of scalar mesons ($f_0, a_0$) in $pp$ reactions is
investigated within a meson-exchange approximation. The relevant
diagrams for the $NN{\to}f_0NN{\to}K\bar{K}NN$ reaction are those
involving pion emission from the nucleons with s-channel
production of the $f_0$-meson and its subsequent decay to
$K\bar{K}$ or $\pi \pi$ (for details the reader is referred to
Ref. \cite{Brat99}). The coupling constant at the $\pi \pi f_0$
vertex is determined from a fit to the reaction $\pi^- p
\rightarrow f_0 n \rightarrow n K^+ K^-$ \cite{Brat99}. A
 form factor at the $f_0\pi\pi$ vertex has to be incorporated
 since both pions
are off their mass-shell; we use the form
\begin{equation}
F_{f_0\pi\pi}(q_1^2,q_2^2)=F_{\pi NN}(q_1^2)F_{\pi NN}(q_2^2) ,
\label{form1}
\end{equation}
where the ${\pi}NN$ form factor is taken as $F(t) = (\Lambda^2 -
m_{\pi}^2)/(\Lambda^2-t)$ with a cut-off parameter
$\Lambda$=1.05~GeV. The form factor~(\ref{form1}) is normalized to
unity at $q_1^2=m_\pi^2$ and $q_2^2=m_\pi^2$, which is consistent
with the kinematical conditions for the determination of the
$f_0\pi\pi$ coupling constant.

\begin{figure}[h]
\psfig{file=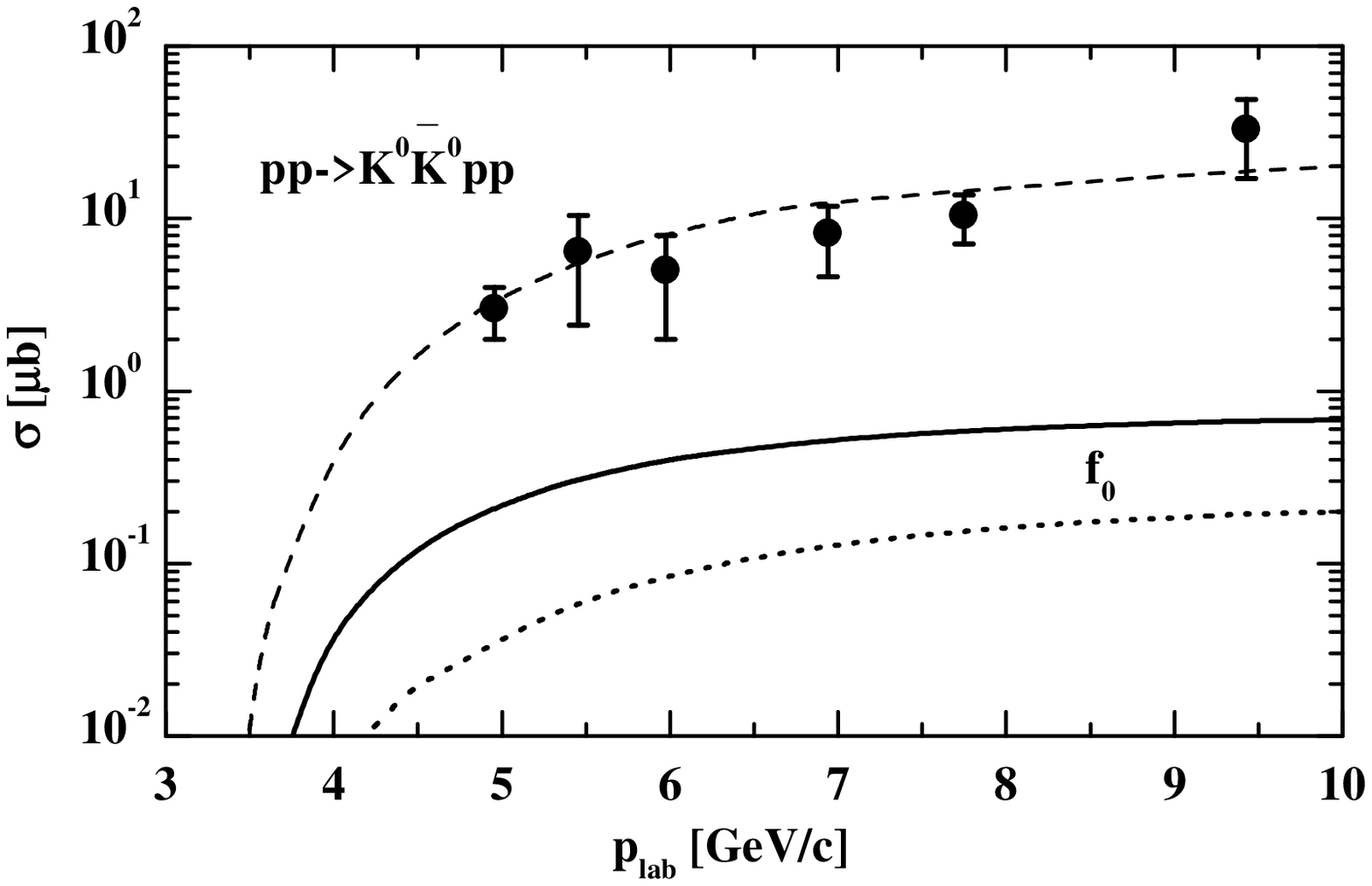,width=5.5cm,height=6cm} \hskip 1cm
\psfig{file=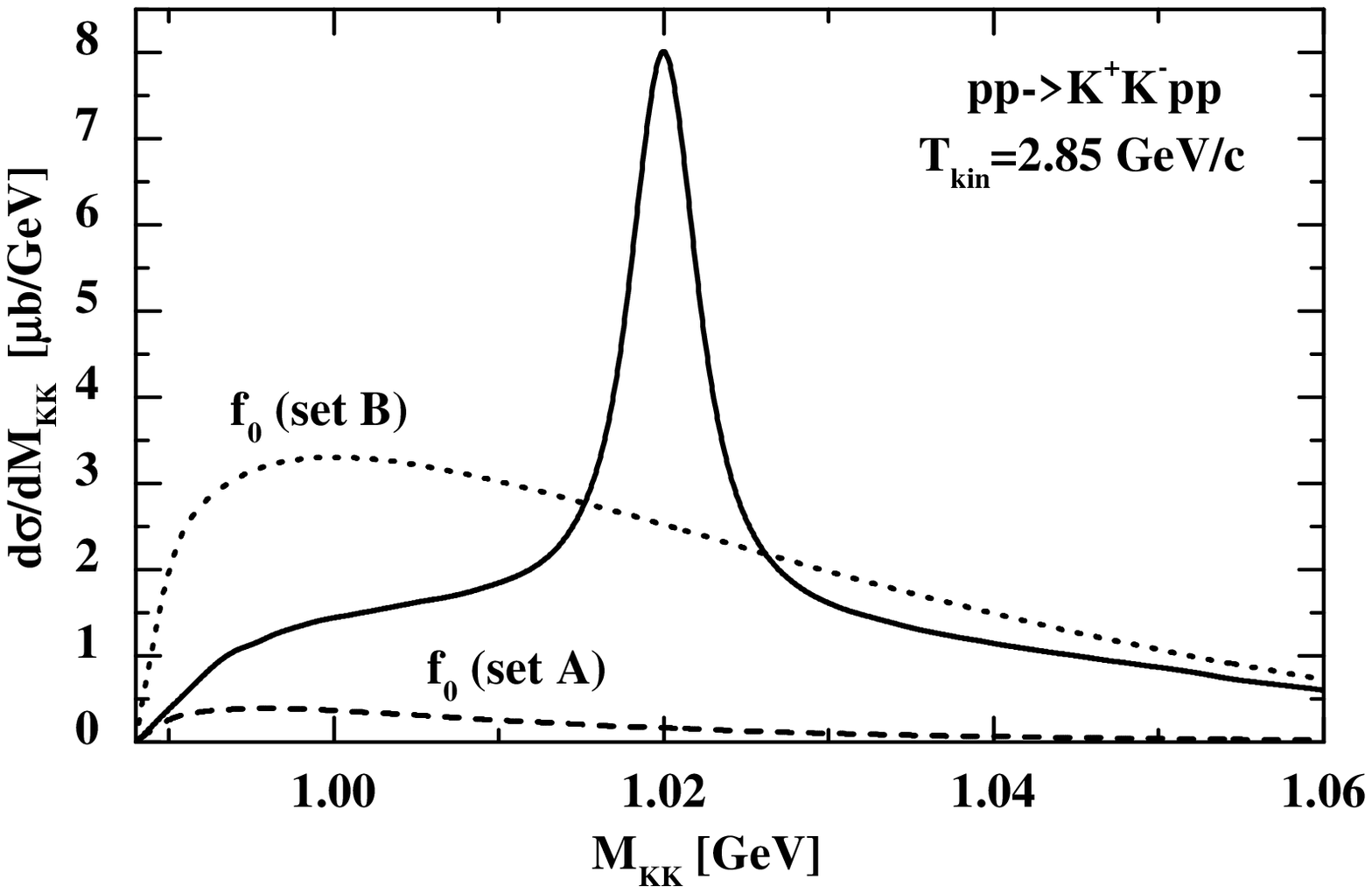,width=5.5cm,height=5.8cm}
\vspace{0.55cm}{\small
\caption{(l.h.s.) The
$pp{\to}f_0pp{\to}K^0\bar{K^0}pp$ cross section calculated with
coupling constants from $set \  A$ with (dotted line) and without
form factor (solid line) at the $f_0\pi\pi$ vertex. The
experimental data for the $pp{\to}K^0\bar{K^0}pp$ reaction are
taken from Ref.~\protect\cite{LB}, while the dashed line shows the
corresponding calculation within the one-boson exchange model from
Ref.~\protect\cite{Sibirtsev1}. (r.h.s.) The $K^+K^-$ invariant
mass distribution from $pp$ collisions at 2.85~GeV. The solid line
shows the calculated contribution from the $\phi$ decays and
$K^+K^-$ background according to Ref. \protect\cite{Sibirtsev1}. The
dashed line shows the contribution from the
$pp{\to}f_0pp{\to}K^+K^-pp$ reaction calculated with constants
from $set \ A$ (lower), while the dotted (upper) line shows the
result obtained with $set \ B$ (see text).}}
\label{Fig7}
\end{figure}

The dotted line in Fig.~\ref{Fig7} (l.h.s.) shows the
$pp{\to}f_0pp{\to}K^0\bar{K^0}pp$ cross section calculated with
the coupling constants $g_{f_0 \pi \pi}$ = 1.49 GeV and g$_{f_0
KK}$ = 0.82 GeV  ($set \ \ A$) and with the form
factor~(\ref{form1}) in comparison to the experimental
data~\cite{LB} for the $pp{\to}K^0\bar{K^0}pp$ reaction. The
dashed line shows the calculations within the pion and kaon
exchange model from Ref.~\cite{Sibirtsev1} for $K\bar{K}$
production. To estimate the maximal $f_0$ production cross section
we neglect the form factor at the $f_0\pi\pi$ vertex and show the
result in terms of the solid line in Fig.~\ref{Fig7} (l.h.s.).
Actually, the contribution from $f_0$ production to the total
$pp{\to}K^0\bar{K^0}pp$ cross section is almost negligible at high
energies. However, a possible way for $f_0$ observation is due to
the low energy part of the $K\bar{K}$ invariant mass spectrum.

We thus calculate the $K^+K^-$ invariant mass spectrum from the
$pp{\to}$ $K^+K^-pp$ reaction at a beam energy of 2.85 GeV, which
corresponds to the kinematical conditions for the DISTO experiment
at SATURNE \cite{DISTO}.  Since at this energy the $\phi$-meson
production becomes possible we include its contribution to the
$K^+K^-$ spectrum. The $pp{\to}{\phi}pp$ total cross section was
taken from Ref.~\cite{Sibirtsev2} and the $K^+K^-$ invariant mass
was distributed according to the Breit-Wigner resonance
prescription with  a full $\phi$-meson width $\Gamma_\phi$ =
4.43~MeV and a branching ratio $Br(\phi\to K^+K^-)=49.1$\%.

The solid line in Fig.~\ref{Fig7} (r.h.s.) shows the $K^+K^-$
invariant mass spectrum for the $pp{\to}{\phi}pp$ reaction as well
as the background spectrum from the $pp{\to}K^+K^-pp$ reaction,
which was calculated as in Ref.~\cite{Sibirtsev1} on the basis of
pion and kaon exchange diagrams. The dashed line indicates the
$K^+K^-$ spectrum calculated with the coupling constants from $set
\ \ A$ and without form factor at the $f_0\pi\pi$ vertex.  For
these coupling constants it is quite obvious that the $f_0$-meson
cannot be directly detected in $pp$ collisions by using the
$K^+K^-$-mode. Note, that when introducing a form
factor~(\ref{form1}) at the $f_0\pi\pi$ vertex the contribution
from $pp{\to}f_0pp{\to}K^+K^-pp$ becomes even smaller.

To test the sensitivity of the model to the $f_0$ parameters we
also performed a calculation with $g_{f_0 \pi \pi}$ = 3.05 GeV and
$g_{f_0 KK}$ = 4.3 GeV ($set \ \ B$) \cite{Brat99} and show the
result in terms of the dotted line in Fig.~\ref{Fig7} (r.h.s.).
Indeed, in this case the $f_0$ contribution is very strong at low
$K^+K^-$ invariant mass, however, the experimental data from DISTO
\cite{Salabura} exclude this possibility.

\subsection{The reaction $pp \rightarrow d a_0^+$}
The missing mass spectrum in the reaction $pp \rightarrow d (
MM)^+$ for deuterons produced at $0^{\circ}$ in the laboratory and
incident momenta of 3.8, 4.5 and 6.3 GeV/c has been measured at
LBL (Berkeley) \cite {Abolins}. It is interesting that apart from
the peaks corresponding to $\pi$ and $\rho$ production, there is a
distinctive structure in the missing mass spectrum at 0.95 GeV$^2$
which was identified as a$_0$ production.

In order to estimate the cross section of the reaction $pp
\rightarrow d a_0^{+}$ at lower momenta -- which are available at
COSY -- the  two-step model (TSM) described by the triangle
diagram is used (\cite{Grishina1,Grishina2}). For the deuteron
wave function we take the parameterization from Ref. \cite
{Lacomb} and neglect the $D$-wave part which gives only a small
contribution compared to the $S$-wave term.

The amplitudes taken into account are: i) the a$_0$ coupling to
two nucleons through the f$_1$(1285) and $\pi $-meson exchanges;
ii) the a$_0$ coupling through $\eta $- and $\pi $-meson
exchanges; iii) the production of a$_0 $-mesons through pion
exchange with $s$- and $u$%
- channel nucleon currents. The coupling constants and cut-off parameters $%
\Lambda _i$ for $\pi $- and $\eta $- meson exchanges are taken
from the Bonn potential model; for the a$_0$- and
f$_1$(1285)-mesons the values from Ref. \cite{Riska} are employed
while the cut-off $\Lambda$  at the nucleon exchange vertex was
considered as a free parameter within the interval 1.2--1.3 GeV.
Within this model the mechanisms i) and ii) are of minor
importance  and the dominant contribution comes from the nucleon
$u$- channel exchange.

The results of the calculations  for the forward differential
cross  section at various beam momenta \cite{Grishina2} match with
the data from \cite{Abolins} for a cut-off $\Lambda$ = 1.3 GeV and
indicate that cross sections of 50 -- 100 nb/sr should be reached
at a bombarding energy of 2.6 GeV, which provides very promising
perspectives for exploring the $a_0$ properties and dynamics at
COSY.

\section{Summary}
In this contribution the production and decay of vector mesons
($\rho, \omega$) in $pA$ reactions at COSY energies has been
studied with particular emphasis on their in-medium spectral
functions. It is found within transport calculations, that
hadronic in-medium decays like $\pi^+\pi^-$ or $\pi^0 \gamma$
might provide complementary information to their dilepton
($e^+e^-$) decays. However, the $\pi^+ \pi^-$ signal from the
$\rho$-meson is strongly distorted by pion rescattering on
nucleons even for light nuclei like $^{12}C$. On the other hand,
the $\omega$-meson Dalitz decay to $\pi^0 \gamma$ appears
promising even for more heavy nuclei since only the neutral pion
may rescatter and  3-photon events ($\pi^0 \rightarrow \gamma
\gamma$) have a very small background at high invariant mass.

Furthermore, the perspectives of scalar meson ($f_0, a_0$)
production in $pp$ reactions have investigated within a
boson-exchange model indicating that the $f_0$-meson might hardly
be detected in the $K \bar{K}$ or $\pi \pi$ decay channels in
these collisions whereas the exclusive channel $pp \rightarrow d
a_0^+$ looks very promising when detecting the deuteron and
analyzing the missing mass spectrum in the range 0.9 -- 1 GeV$^2$.

\end{document}